\documentstyle[aps,multicol,epsfig]{revtex}                                                                                                                                                                                                                                                                                                                                                                                                                                                                                                                                                                                                                                                                                                                                                                                                                                                                                                                                                                                                                                                                                                                                                                                                                                                                                    \begin{document}
\title{A low-temperature dynamic mode scanning force microscope
operating in high magnetic fields}
\author{J. Rychen, T. Ihn, P. Studerus, A. Herrmann, K. Ensslin}
\address{Laboratory of Solid State Physics, ETH H\"onggerberg,
CH-8093 Z\"urich, Switzerland}
\date{\today}
\maketitle

\begin{abstract}
A scanning force microscope was implemented operating at
temperatures below 4.2K and in magnetic fields up to 8T. Piezoelectric
quartz tuning forks were employed for non optical tip-sample distance
control in the dynamic operation mode. Fast response was achieved by
using a phase-locked loop for driving the mechanical oscillator.
Possible applications of this setup for various scanning probe techniques
are discussed.
\end{abstract}


\begin{multicols}{2}
\narrowtext

\section{Introduction}
Low temperature scanning force microscopes (SFMs) are desirable
for many applications in physical research. To our knowledge, the
first prototypes have become commercially available only recently,
reaching base temperatures around 5K under UHV-conditions. Lower
temperatures can be reached with only a few home-built microscopes
in non-UHV setups\cite{pelekhov,eriksson,crook}.

Especially in cases where the sample itself is sensitive to light 
a non-optical detection method is needed in order to keep the  
probe-sample distance fixed. Such a method has the additional 
advantage that it leads to less involved setups. Piezoresistive 
cantilevers have been proposed \cite{tortonese} and applied for 
low temperature scanning force microscopy\cite{eriksson,crook}. 
However, these cantilevers are not easily available. Recently, 
piezoelectric quartz tuning forks have been employed in a scanning 
near-field optical microscope designed for operation at low 
temperatures\cite{karrai}. In this setup, an optical fibre is 
glued along one prong of the tuning fork which is mechanically 
excited to oscillate. The tuning fork is used as a friction-force 
sensor. The advantages of these piezoelectric sensors are the 
availability, the low cost and the high quality factors. These 
tuning forks have been successfully employed for atomic force 
microscopy\cite{edwards}, scanning near field optical 
microscopy\cite{karrai,ruiter,atia,salvi,tsai}, magnetic force 
microscopy \cite{todorovic} and acoustic near field microscopy 
\cite{steinke} at room temperature. In Refs. 
\onlinecite{karrai,ruiter,salvi,tsai,todorovic} the SFM operation 
was just used to keep the tip-sample distance fixed while an 
additional nanosensor measures the physical quantity of interest. 

In this paper we describe the implementation of a tuning fork sensor
suitable for SFM-imaging at temperatures below 4.2K and in high
magnetic fields.

\section{Microscope Overview}
\begin{figure}
\noindent\centerline{\epsfig{file=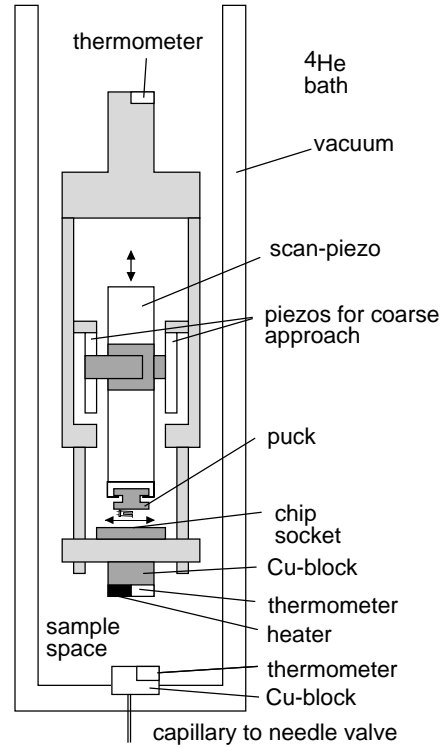,width=0.7\linewidth}}
\caption{Schematic microscope overview} \label{fig1}
\end{figure}

The microscope is based on an Oxford Instruments made commercial 
cryo-SXM being significantly modified for operation as an SFM for 
probing semiconductor devices. Fig. \ref{fig1} shows the schematic 
setup. The microscope head which is made of nonmagnetic material 
(mainly Ti and CuBe) is mounted at the end of a sample rod which 
can be suspended in the sample space of a 1.7K variable 
temperature insert (VTI). The VTI is part of a standard $^4$He 
cryostat with a superconducting magnet producing magnetic fields 
up to 8T. The temperature can be controlled by setting the gas 
flow with a needle valve and controlling the heater power. The 
head can be operated under ambient conditions  or in the cryostat 
in $^4$He gas at temperatures between 300K and 2.2K and at 
pressures of typically a few millibars. Below 2.2K the helium 
becomes superfluid and the operation becomes difficult for reasons 
discussed below. 

Effective insulation of the cryostat from building vibrations is
achieved by suspending it from appropriate rubber-ropes leading to
a resonance frequency of about 1Hz for the whole system. The 
$^4$He pumping line and the He-recovery line are plastic pipes 
which are suspended together with all electrical connections to 
the cryostat from additional elastic ropes. Vibrations from the 
$^4$He pump are effectively damped by guiding the pumping line 
through a box of sand. Microphony of the cryostat is reduced using 
a rubber mat which is tied around the cryostat body. Additional 
vibration isolation between cryostat and sample rod is implemented 
using a specially designed top flange with passive damping 
elements and adjusting screws to prevent mechanical contact 
between the microscope and the VTI tube. 

The microscope head allows coarse tip-sample approach using a 
slip-stick drive moving the scan-piezo up or down. A puck mounted 
at the end of the scan-piezo can be laterally positioned in a 
similar fashion. In contrast to the commercial design of the head, 
in our system it serves as the platform where the SFM sensor is 
mounted. Samples are mounted in chip carriers which can be easily 
plugged into a 32-pin chip socket. The socket is mounted on a 
copper block which incorporates the sample heater and the sample 
thermometer. The heater allows us to evaporate the water film from 
the sample surface before cooldown. It further gives us the 
possibility to keep the sample warmer than its surroundings during 
the cooling process in order to avoid freezing contaminations on 
the sample surface. The sample mount unit can be used for standard 
magnetotransport measurements independent of the SFM operation. 
For detection of temperature gradients, additional thermometers are 
located at the bottom of the VTI and at the top of the microscope 
head. The scanning unit is a five electrode tube scanner of 50.8mm 
length and outer diameter of 12.7mm. With a maximum (bipolar) 
voltage of 230V it gives a lateral scan range of 52.2$\mu$m in 
$x$-$y$ direction and a $z$-range of 5$\mu$m at a temperature of 
290K. At 4.2K the lateral range is 8.8$\mu$m and the $z$-range is 
0.85$\mu$m. The TOPS3 control electronics by Oxford Instruments 
consist of a digital feedback loop which can be switched from 
logarithmic amplification for STM operation to linear 
amplification for the SFM mode. Tip approach and data acquisition 
are computer controlled. 

\section{Piezoelectric tuning forks}
\begin{figure}
\noindent\epsfig{file=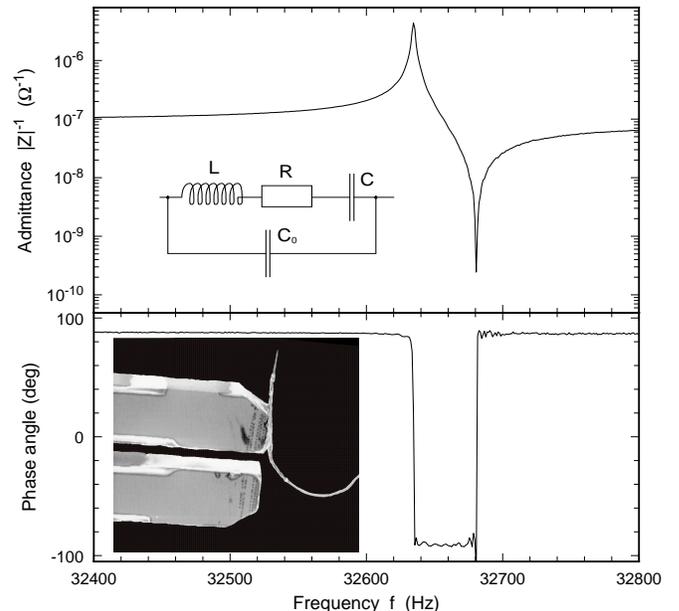,width=\linewidth}
\caption{Typical resonance curve for a tuning fork at 4K in 
Vacuum. Admittance and phase versus driving frequency are shown. 
The top inset shows the tuning fork equivalent circuit. In the 
lower inset a scanning electron micrograph of a tuning fork with a 
13$\mu\mathrm{m}$ W-wire glued to the end of one prong is shown.} 
\label{fig2} 
\end{figure}

In order to implement cheap and easy-to-make dynamic SFM operation 
at cryogenic temperatures we decided to avoid optical cantilever 
deflection detection and use piezoelectric tuning forks instead. 
Initially, quartz tuning forks have been developed for the 
realization of very small and stable oscillators in watches. Due 
to their use in industry they are cheap and easily available. Most 
of them have a (lowest) resonance frequency $f_{0}=2^{15}$Hz and 
quality-factors in vacuum between $Q=20000$ and $100000$. In 
scanning probe microscopy they have recently found new 
applications as piezoelectric 
sensors\cite{karrai,edwards,ruiter,atia,salvi,tsai,todorovic,steinke}. 
The high $Q$-values of tuning forks offer the advantage that 
dynamic mode operation under ambient conditions or in liquids is 
possible. However, low-temperature operation in the dynamic 
SFM-mode has not been demonstrated before. In our system the 
tuning fork is driven by an AC-voltage from a Yokogawa function 
generator FG320.  This voltage is scaled down by a factor of 1000 
with a voltage divider residing close to the microscope head, 
which leads to typical excitation amplitudes of 0.1-10mV. We 
directly measure the admittance of the tuning fork with a 
home-made current-voltage converter with a gain of 1V/$\mu$A 
connected in series to the tuning fork. The mechanical tip 
amplitude depends linearly on the measured current and is 
typically between 1-100nm\cite{karrai,rychen}. Due to the high 
oscillator quality the power loss of the tuning fork at resonance 
can be far less than $10\textrm{nW}$, which makes it ideal for the 
use in low cooling power cryostats. A typical resonance curve and 
a corresponding equivalent circuit is shown in Fig. \ref{fig2}. 
The mechanical resonator is modeled by the $LRC$-branch in the 
equivalent circuit which leads to the admittance maximum. The 
capacitance between the electrodes described by $C_0$ leads to the 
asymmetry of the resonance curve with a minimum above the 
resonance frequency\cite{rychen}. 

A tip made of W or PtIr wire with typical diameter of 20$\mu$m is
glued to one of the tuning fork arms in parallel to the direction of
the vibrational motion of the arm (see inset of Fig. \ref{fig2}). 
The tip can be contacted separately. After mounting the tuning 
fork with the tip on the scanning head the tip is 
electrochemically etched to achieve tip radii of less than 30nm. 
After the preparation of a tip the resonance frequency of the 
tuning fork is shifted by about 100Hz to lower frequencies and the 
$Q$-value is lowered but can be kept above a value of 20000 in 
vacuum. 

The tip prepared on the tuning fork can be used as a tunnelling 
tip (STM-mode) or as the tip in dynamic SFM operation mode without 
any modifications on the scanning head. This gives simultaneous 
access to two complementary imaging modes at low temperatures 
without the need of a tip exchange. The high stiffness of the 
tuning fork arms (static spring constant up to 2$\mu$N/nm) is 
sufficient to guarantee stable tunnelling conditions. 

\section{Phase-locked loop}
The principle of dynamic mode SFM operation of the tuning fork is 
the same as for normal cantilevers. The elastic interaction of the 
tip with the sample surface will shift the resonance frequency via 
the presence of force gradients. Inelastic tip-sample interactions 
will mainly alter the $Q$-value of the oscillator. With our setup 
we can use both quantities to control the z-piezo via the feedback 
loop, i.e. we can either keep the resonance frequency or the 
dissipated power constant during a scan. 

\begin{figure}
\noindent\epsfig{file=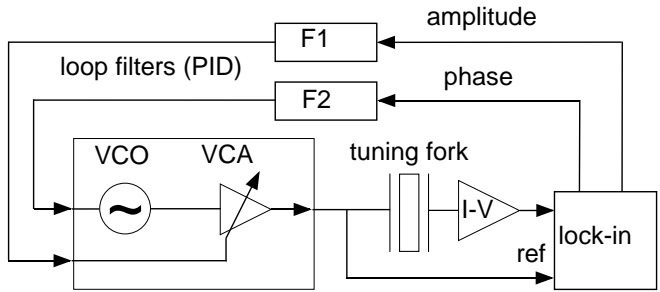,width=\linewidth} \vspace{3mm} 
\caption{Operation principle of the phase-locked loop. A voltage 
controlled oscillator (vco) followed by a voltage controlled 
amplifier (vca) is driving the tuning fork. The phase between the 
driving signal and the measured current is detected by a lock-in 
amplifier and its output is used to control the frequency and 
amplitude.} \label{fig3} 
\end{figure}

If driven at constant frequency, the high $Q$-values of tuning
forks lead to a very slow response of the oscillation amplitude or
phase on a steplike change in the tip-sample interaction on a time
scale $Q/f_{0}$. The resulting limitation in bandwidth is
undesirable since it makes SFM imaging very slow. A significant
increase in the bandwidth can be achived with the use of a
phase-locked loop\cite{edwards,atia,albrecht}. Figure \ref{fig3} 
shows the operation principle. The measured current through the 
tuning fork is analyzed electronically and its amplitude and phase 
(relative to the driving voltage) are determined. Both signals are 
fed back into the oscillator via two separate PID-components. The 
phase signal is used to modulate the excitation frequency and 
allows locking on a fixed value of the phase. In contrast to the 
detection of amplitude or phase changes at fixed driving 
frequency, the phase-locked loop gives a much faster response on 
the time scale needed for the determination of the signal phase, 
which is typically $3/f_{0}\approx 100\mathrm{\mu s}$ independent 
of the $Q$-value. This scheme allows us to achieve a bandwidth 
significantly larger than that of the $z$-feedback loop. The 
latter is limited by the lowest mecanical resonance frequency of 
the scan piezo at around 1kHz.  

When we prepare for scanning, we first measure the frequency 
response of the tuning fork to determine the actual resonance 
frequency $f_{0}$. The phase-locked loop can then be set up with 
$f_{0}$ as the carrier frequency. The frequency of the function 
generator is controlled with a sensitivity set to $100$ - 
$500\mathrm{mHz}/V$. The noise in the frequency is then of the 
order of $100\mu\mathrm{Hz}/\sqrt{Hz}$ in the range from 1Hz to 
1kHz corresponding to 10 mHz peak-to-peak.  

As an additional refinement we can use the measured amplitude of
the tuning fork oscillation to feed it back into the amplitude
modulation input of the oscillator. This additional
feedback keeps the amplitude of the tuning fork oscillation at a
fixed value by adjusting the amplitude of the excitation voltage. 

\section{SFM operation}
As the error signal for the feedback loop controlling the tip 
sample distance we use either the voltage proportional to the 
frequency shift (frequency control mode) or the change in driving 
voltage needed to keep the oscillation amplitude constant 
(amplitude control mode). Due to the size of our scan piezo its 
mechanical resonances allow a bandwidth of only 1kHz for the 
$z$-control, i.e.  this is the decisive factor limiting the 
overall bandwidth of our system. We therefore reach typical scan 
speeds of up to 10$\mu$m/s. 

Before the microscope is inserted into the cryostat, the sample is 
heated above 100$^\circ$C and the VTI is heated close to room 
temperature. After inserting the microscope we pump the VTI and 
cool down at a rate of 3K/min. The sample is kept 50K above the 
temperature of the gas flow in order to prevent freezing 
contaminations on the sample surface. For optimum operation a 
stable temperature gradient along the sample rod has to be 
maintained. The pressure, which strongly affects the resonance 
frequency, has to be stabilized below the vapor pressure of $^4$He 
in order to prevent liquid helium to enter the VTI. Operation in 
normal liquids (e.g. $^4$He or $^3$He) is possible but the 
$Q$-value and thereby the sensitivity is reduced by a factor of 
four. Below 2.2K, where $^4$He becomes superfluid, in our cryostat 
the resonance frequency becomes unstable, presumably due to 
thermodynamic instabilities. These problems, however, can be 
solved by operating the microscope in a vacuum beaker.         

\begin{figure}
\noindent\epsfig{file=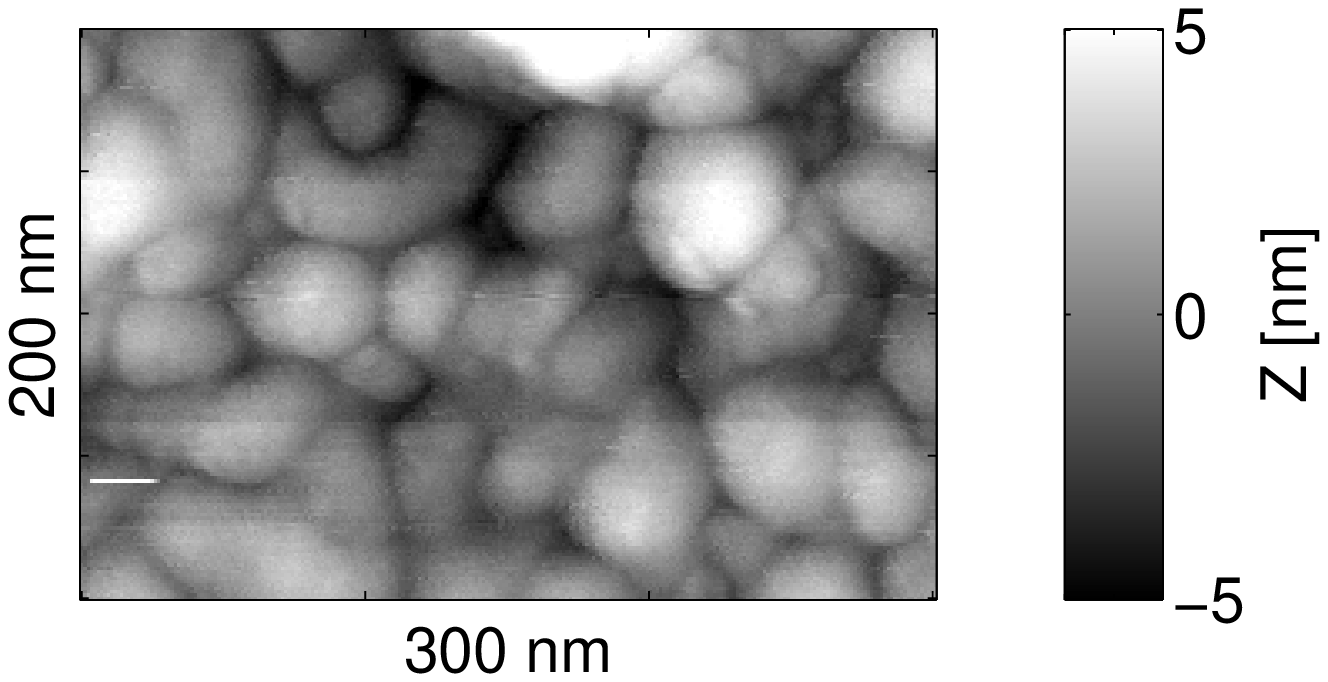,width=\linewidth}
\epsfig{file=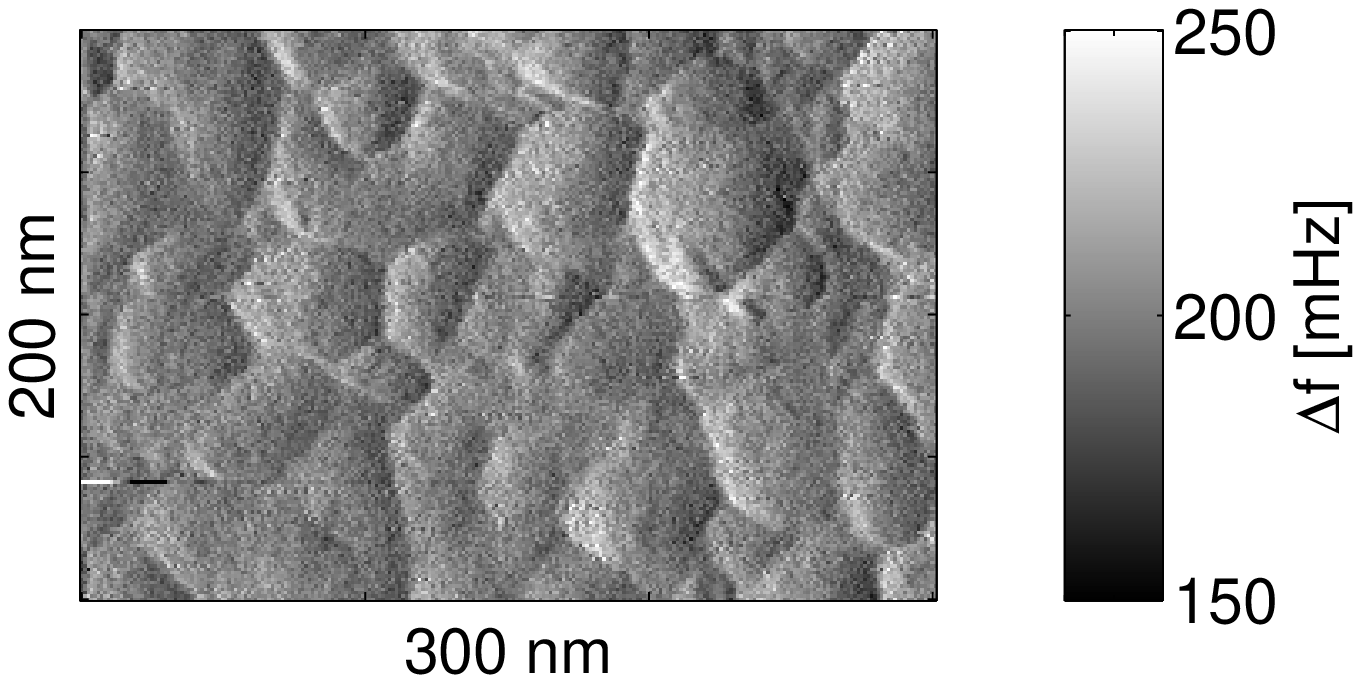,width=\linewidth}
\caption{Surface of a 150nm gold film evaporated onto a glass 
substrate, which shows grains of 30nm size ($T=2.5$K). Top image: 
topography, bottom image: frequency shift} \label{fig4} 
\end{figure}

In the following we show two examples of SFM-operation at low 
temepratures. Fig. \ref{fig4} shows the image of the surface of a 
150nm gold film evaporated on a glass substrate. Small grains with 
a typical size of 30nm are resolved. The lateral resolution is 
better than 20nm, the resolution in $z$-direction is better than 
1\AA. The upper image shows the topography and the lower image is 
the frequency shift, i.e. the error signal for the $z$-feedback.   

\begin{figure}
\noindent\epsfig{file=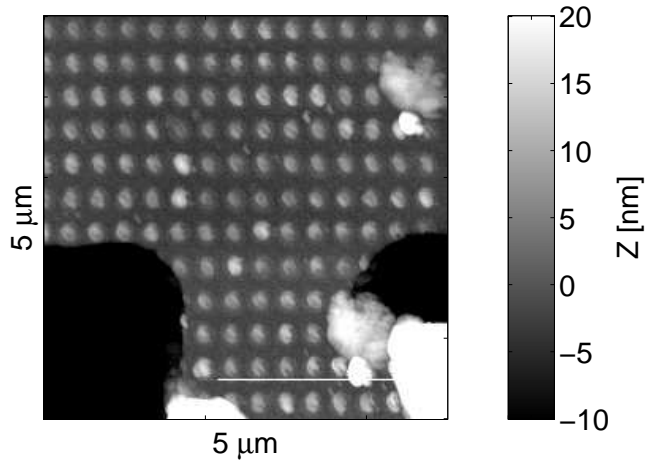,width=\linewidth}
\caption{Square lattice of oxide dots on top of a GaAs/AlGaAs
Hall bar structure imaged at 2.5K.}
\label{fig5} 
\end{figure}

If one sets out to investigate semiconductor nanostructures, 
locating the structure of interest at low
temperature where 
the scan range is only a few microns, is a significant problem. 
With the
lateral coarse tip positioning we were able to find a
10$\mu$m$\times 10\mu$m spot on top of a
GaAs/AlGaAs-heterostructure where a Hall-bar device had been
fabricated by photolithography techniques. On top of a part of the
Hall-bar a square lattice of oxide dots with a period of 400nm had
been written by AFM-lithography\cite{held}. Fig. \ref{fig5} shows 
an image of this area of the sample taken at a temperature of 
2.5K.  

Operation of the head in a magnetic field of 8T shifts the scanned 
area by about 0.5$\mu$m. It is therefore possible to scan at 
different fixed magnetic fields without loosing the structure of 
interest. Spectroscopy at a fixed point of the surface with 
sweeping magnetic field is difficult. The resonance frequency of 
the tuning fork is not found to shift significantly when the 
magnetic field is altered. 

The design of our cryo-SFM, especially the low power dissipation 
of the tuning forks, makes it an ideal system for imaging at even 
lower temperatures in a $^3$He-cryostat or even a dilution 
refrigerator. In spite of the high stiffness of the tuning fork 
sensor compared to conventional SFM-cantilevers high sensitivity 
is achieved due to the large $Q$-values of the mechanical 
oscillator. The high stiffness allows the preparation of probes 
without a strong impairment of oscillator performance. 
It may offer the option in the future to fabricate other sensors, 
e.g. semiconductor chips, instead of or in addition to tunnelling 
tips at the end of one tuning fork arm. A straightforward 
continuation of the development of our system leads to the 
application of the scanned gate technique,  Kelvin force 
microscopy and scanning capacitance microscopy. 

\section{Conclusion}
In conclusion, we have implemented a low-cost dynamic mode
cryo-SFM for operation down to temperatures below 4.2K and in
magnetic fields up to 8T. Due to the utilization of piezoelectric
quartz tuning forks there is no need for an optical cantilever
deflection detection. The unit can be operated in STM-mode or
SFM-mode without tip exchange. High bandwith is achieved with a
phase-locked loop which controls the driving frequency of the
tuning fork. Further development beyond the SFM-operation at low
temperatures is feasible.

\acknowledgments

The authors would like to thank H.~Hug and K.~Karrai for valuable 
discussions. Financial support by the Eidgen\"ossische Technische 
Hochschule is acknowledged.

\end{multicols}
\end{document}